# Longitudinal Single Bunch Instability Study on BEPCII[*]

WANG Dou（王逗）[1)], LI Yong（李勇）, DUAN Zhe（段哲）, WANG Na（王娜）,

WANG Li（王理）, WANG Lin（汪林）, GAO Jie（高杰）

Institute of High Energy Physics, CAS Beijing 100049, China

**Abstract**: In order to study the single bunch longitudinal instability in BEPCII, experiments on the positron ring (BPR) for the bunch lengthening phenomenon were made. By analyzing the experimental data based on Gao's theory, the longitudinal loss factor for the bunch are obtained. Also, the total wake potential and the beam current threshold are estimated.

## 1. Introduction

The bunch lengthening phenomenon in an electron storage ring was first observed in ACO [1] at Orsay and later in other machines, where accompanying bunch lengthening, one finds an increase in the single bunch energy spread with a more or less appreciable threshold current. The first empirical bunch lengthening formula [1] found in ACO is expressed as:

$$\sigma_\tau^2(ns) = \sigma_{\tau 0}^2(ns)\left(1 + 2\times 10^{-3}\frac{I_b(mA)}{E^4(GeV)\sigma_\tau(ns)}\right), \quad (1)$$

where the energy dependence $E^4(GeV)$ in eq. (1) was corrected to be $E^3(GeV)$ in the later theoretical works, such as refs. [2] and [3]. Afterwards, the theories largely used to describe the bunch lengthening phenomenon are potential well distortion (PWD) and microwave instability [4] theories. Additionally, A. W. Chao and J. Gareyte suggested a universal scaling law which is independent of the model chosen to account for bunch lengthening [5]. The "Chao-Gareyte scaling law" has been a very important tool for the accelerator physicists and the machine designers. However, the problems in this field seems still not cleaned up, just as stated by Chao [6] and Gareyte [7], the field of bunch lengthening is still an open area and more theoretical works should be devoted in the classical problem.

In order to study the single bunch longitudinal instability of BEPCII, the relation of bunch length and beam current for BPR was measured in 2011 April and June respectively. The experimental results were analyzed according to the method in ref. [8] and the conclusion after calculations was that the impedance got from bunch lengthening measurement is about 3 times larger than that of impedance budget [9], that is similar as the observation in ref. [10]: "The impedance values implied by our analysis are much larger than expected: L is about a factor of 3 larger than earlier calculations, and R is about a factor of 10 larger, a result that is especially puzzling.".

In this paper, we try to analyze the experimental results by Gao's theories [11-13] which are based on the concept of collective random excitations and can predict the bunch length and energy spread in the whole current range. First, we give a brief review of corresponding theories. Then we make an explanation about how to solve the experiment data and how to calculate the longitudinal loss factor by these theories. Also we make estimation for the wake potential of the whole ring and give a prediction for the bunch lengthening and energy spread increase according to the wake potential. Finally, the threshold beam current is estimated.

## 2. Review of theory

The most popular theory for the single bunch longitudinal collective effects when the beam current is lower than the microwave instability threshold is the potential well distortion theory, which is expressed as follows [2,3]:

---
[*] Supported by the National Foundation of Natural Sciences Contract 11175192.
1） Corresponding author at: +86-10-88236743; E-mail address: wangdou@ihep.ac.cn (D. Wang).

$$R_z^2 = 1 + \frac{C_{PWD} I_b}{R_z}, \quad (2)$$

$$R_\varepsilon = 1, \quad (3)$$

where $R_z = \frac{\sigma_z}{\sigma_{z0}}$, $R_\varepsilon = \frac{\sigma_\varepsilon}{\sigma_{\varepsilon 0}}$, and $C_{PWD}$ is a coefficient corresponding to the potential well distortion effect. In this theory, the beam energy spread will keep constant. Only as the beam current is increased larger than the threshold, the energy spread will increase and the mechanism which determine the bunch lengthening switch to the microwave instability effect.

What is very important to realize is that the phenomenon of the longitudinal instabilities in an electron storage ring is quite different from that in a proton storage ring where the synchrotron radiation does not play a dominant role [14]. In electron storage rings, the particles are heated by random quantum radiation excitations otherwise the particles are cold in proton machines. So each particle in electron machines is random and its trace is unpredictable just like the gas molecular whose behaviour is governed by the Fokker-Planck equation. In fact, the word "micromave instability" came from the proton machine. In the electron machines, the beam energy spread always increases and we can not find an obvious turning point for the micromave instability threshold. The energy spread measurement in ATF

For the first time, Gao introduced the concept of collective random excitations by the longitudinal wake potential and improved the potential well distortion theory to [11]

$$R_z^2 = 1 + \frac{C_{PWD} I_b}{R_z^{1.5}} + \frac{\chi [R_{av} R I_b k(\sigma_{z0})]^2}{\gamma^7 R_z^{2\varsigma}}, \quad (4)$$

$$R_\varepsilon^2 = 1 + \frac{\chi [R_{av} R I_b k(\sigma_{z0})]^2}{\gamma^7 R_z^{2\varsigma}}, \quad (5)$$

$$\chi = \frac{576 \pi^2 \varepsilon_0}{55 \sqrt{3} \hbar c^3}, \quad (6)$$

$$k(\sigma_z) = k(\sigma_{z0}) / \left(\frac{\sigma_z}{\sigma_{z0}}\right)^\varsigma, \quad (7)$$

where $R_{av}$ is the average radius of the ring, $R$ is the local bending radius, $\gamma$ is the normalized particle energy, $k(\sigma_{z0})$ is the longitudinal loss factor at the nature bunch length (zero current) and $\varsigma$ comes from the scaling law (7) (each machine has its own $\varsigma$ and $\varsigma \approx 1.21$ for SPEAR scaling law). Reference [11] shows a quite good agreement between this new theory and the experimental results.

Later on, an empirical expression of $C_{PWD}$ was given based on the reference [11] and finally Gao's theories for bunch lengthening was summarized as follows [12] (SPEAR scaling law $\varsigma \approx 1.21$ is used):

$$R_z^2 = 1 + \frac{\sqrt{2} A I_b}{R_z^{1.21}} + \frac{(A I_b)^2}{R_z^{2.42}}, \quad (8)$$

$$R_\varepsilon^2 = 1 + \frac{(A I_b)^2}{R_z^{2.42}}, \quad (9)$$

$$A = \frac{\sqrt{\chi} R_{av} R k(\sigma_{z0})}{\gamma^{3.5}}. \quad (10)$$

These empirical formulae give our precious convenience that we can get the longitudinal loss factor if we fit the coefficient $A$ from the measurements of bunch lengthening.

### 3. Experimental results and analyses

#### 3.1. Bunch lengthening measurement and broad-band impedance estimation

The bunch length in BPR of BEPCII was measured by the cooperation of accelerator physics group and beam instrumentation group with streak camera [9]. Each time, a single bunch was injected into the BPR and the bunch length at different beam current was measured. During the measurements, we found that the scan scales and the gain of streak camera have strong influences on the results. In this paper, we choose the experiment results of April 26th (The beam parameters on this day are closest to that for machine design so that we can make a comparison with the impedance budget.) and firstly fit these data to get a roughly broad-band impedance using the "Chao-Gareyte scaling law".

In Fig. 1, we have plotted $\sigma_z$ as a function of the beam current. It shows that $\sigma_z \propto I^{1/6.28}$ which means a=4.28, or in the frequency range from a few GHz to 10 GHz ($c/2\pi\sigma_z$=3.2GHz), the longitudinal impedance increases as $Z(\omega) \propto \omega^{4.28}$.

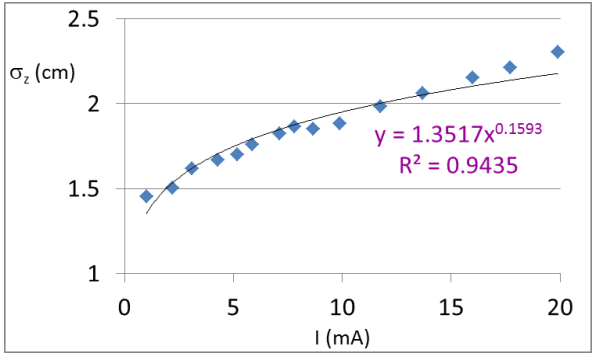

Figure 1: Bunch length vs. beam current for 1.5 MV rf voltage and 1.89 GeV beam energy.

### 3.2. Longitudinal loss factor estimation

Then we use Gao theory to get the longitudinal loss factor. The relative bunch length at different beam current is shown in Figure 2 and the corresponding beam parameters of BPR for this measurement are in Table 1.

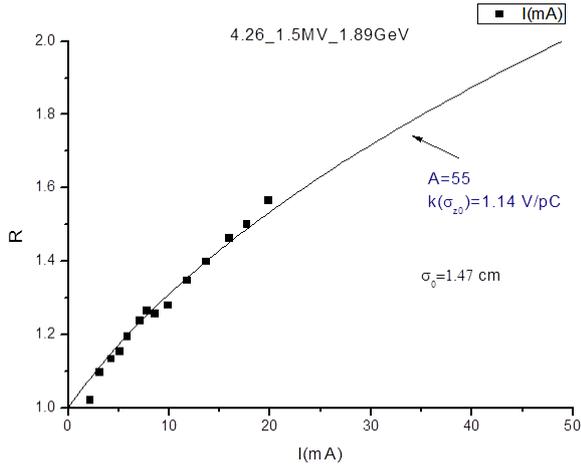

Figure 2: Relative bunch length vs. beam current for 1.5 MV rf voltage and 1.89 GeV beam energy (dots: experimental results, line: fit by using eq. (8)).

Table 1: Main beam parameters of BPR corresponding to the measurement of April 26th

| Parameters | BPR(4.26) |
|---|---|
| $E_0$ (GeV) | 1.89 |
| $V_{rf}$ (MV) | 1.5 |
| $\nu_{s0}$ | 0.0319 |
| $\alpha_p$ | 0.0241 |
| $\sigma_{z0}$ (cm) | 1.47 |
| $\sigma_\varepsilon$ | 5.16×10$^{-4}$ |
| $N_b$ | 1 |
| $I_b$ (mA) | 1~20 |
| $\phi_{s0}$ | 175° |

The experimental results are fit by using the equation (8) (see Figure 2). From the fitting, we get the coefficient A=55, and then according to equation (11), the longitudinal loss factor ($k(\sigma_{z0}) = 1.14 V/pc$) is calculated. The impedance budget of longitudinal loss factor at 1.5 cm bunch length from BEPCII design report (edition II) is 1.76 V/pC [15]. Considering we have removed the profiles since 2008, the total budget should be smaller than 1.76 V/pC. So our estimation for longitudinal loss factor basically agrees with the impedance budget.

### 3.3. Wake potential estimation for BPR

According to the impedance budget of BEPCII, the total inductance of BPR is about 69nH [9], which have included the longitudinal feedback kicker and the new injection kickers. With the inductance ($L$=69nH) from the impedance budget and the loss factor ($k(\sigma_{z0}) = 1.14 V/pc$) from the experiment, we can estimate the wake potential for the whole ring of BPR by using Gao's theory [13] which is expressed as follows:

$$W_z(z) = -ak(\sigma_z)\exp(-\frac{2z^2}{7\sigma_z^2})\cos[(1+\frac{2}{\pi}arctg(arctg(\frac{Z_i}{2Z_r})))\frac{z}{\sqrt{3}\sigma_z}+arctg(\frac{Z_i}{2Z_r})]$$

$$= Ak(\sigma_z)+Bk(\sigma_z)\frac{z}{\sigma_z}+Ck(\sigma_z)(\frac{z}{\sigma_z})^2+O(z^3) \quad (11)$$

$$(A=-\frac{a}{\sqrt{1+\left(\frac{Z_i}{2Z_r}\right)^2}}, \quad B=\frac{0.289aZ_i(1+0.637arctg(arctg(\frac{Z_i}{2Z_r})))}{Z_r\sqrt{1+\left(\frac{Z_i}{2Z_r}\right)^2}}, \quad C=\frac{a}{\sqrt{1+\left(\frac{Z_i}{2Z_r}\right)^2}}\left(\frac{2}{7}+\frac{\left(1+0.637arctg(arctg(\frac{Z_i}{2Z_r}))\right)^2}{6}\right))$$

where $a=2.23$, $Z_i=2\pi L/T_0$, $Z_r=k(\sigma_z)T_b^2/T_0$, $T_0=2\pi R_{av}/c$, $T_b=3\sigma_z/c$, $\sigma_z$ is the bunch length, c is the velocity of light, and $z=0$ corresponds to the centre of the bunch. Figure 3 shows the predicted wake potential of the whole ring for BEPCII BPR based on eq. (11).

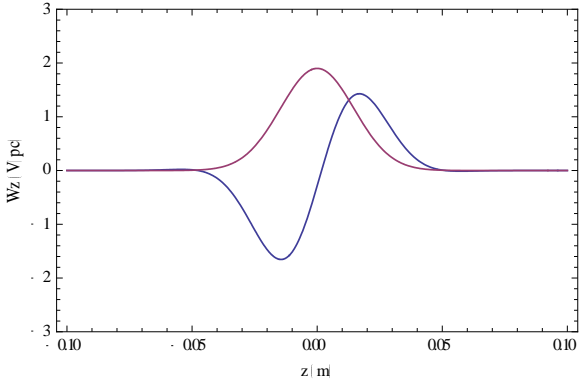

Figure 3: Analytical wake potential (blue line) of the whole ring for BEPCII BPR based on eq. (11), with $\sigma_{z0}=0.0147$m, $L=69$nH, and $k(\sigma_{z0})=1.14$V/pc. The red line shows the Gaussian bunch shape with arbitrary units.

With the wake potential and its Taylor expansion, the bunch length, the energy spread and the current threshold can be expressed by another way [13],

$$R_z^2 = 1+\frac{2eBk(\sigma_{z0})R\sigma_{z0}I_b}{m_0c^3\alpha_p C_q\gamma^3 R_z^{1.21}}+\frac{\chi[R_{av}RBk(\sigma_{z0})I_b]^2}{\gamma^7 R_z^{2.42}}, \quad (12)$$

$$R_\varepsilon^2 = 1+\frac{\chi[R_{av}RBk(\sigma_{z0})I_b]^2}{\gamma^7 R_z^{2.42}}, \quad (13)$$

$$I_{e,th} = \sqrt{\frac{2}{3}}\frac{\sigma_{z0}V_{rf}\cos(\phi_{s0})}{Ck(\sigma_{z0})T_0^2 f_{s0}\lambda_{rf}}, \quad (14)$$

where $B$ and $C$ are the second and third-order coefficient of Taylor expansion of wake potential, $I_{e,th}$ is the so called phase instability threshold above which the particles will execute stochastic motions, $V_{rf}$ is the peak rf voltage, $\phi_{s0}$ is the synchronous phase, $T_0$ is the revolution period, $f_{s0}$ is the synchrotron oscillation frequency and $\lambda_{rf}$ is the wave length of rf field.

We can see that the linear coefficient B of wake potential will determine the bunch lengthening and energy spread increase. The prediction of bunch lengthening and energy spread increasing for BEPCII BPR is shown in Figure 4. Otherwise the nonlinear coefficient C will determine the current threshold of instability refer to longitudinal stochastic motion.

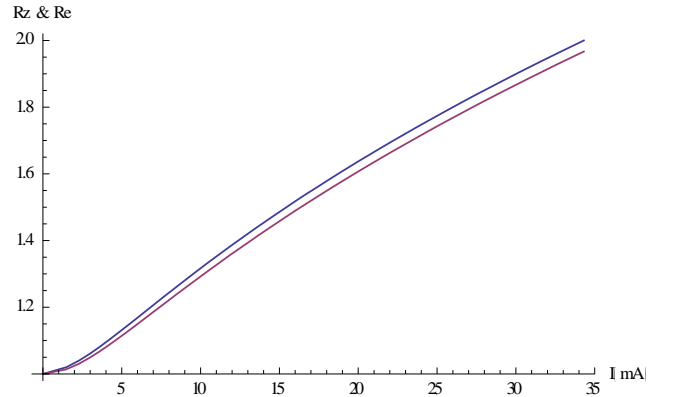

Figure 4: Estimation of bunch lengthening and energy spread increasing for BEPCII BPR (blue line: bunch lengthening, red line: energy spread.) based on eq. (12) and eq. (13), with $\sigma_{z0}=0.0147$m, $L=69$nH, $k(\sigma_{z0})=1.14$V/pc, $B=2.07$ and $C=0.18$.

From Figure 4, we can see that the analytical prediction of bunch lengthening agrees well with the experimental results in Figure 2. So to a great extent, it proves that the value of loss factor ($k(\sigma_{z0})=1.14$V/pc) from experiment and the inductance ($L=69$nH) from the impedance budget are reasonable.

### 3.4. Beam current threshold estimation for BPR

Knowing the nonlinear coefficient C of wake potential, we can also predict the threshold for beam current from eq. (14). By calculation, the threshold is 5.69 mA with $L$=69nH, $k(\sigma_{z0})$=1.14V/pc and C=0.18.

In addition, we made a test by increasing the inductance and the loss factor respectively. If we increase the loss factor to 3V/pc and keeping the 69nH inductance, the threshold current dropped to 0.88mA quickly. However the threshold rises to 23.3mA by increasing the inductance to 287nH (The measured inductance of BPR is 287nH in reference [9].). It seems that larger loss factor can aggravate the longitudinal instability but larger inductance may have positive effect. This test reminds us that the loss factor which is the resistive part of the impedance has significant effect on the bunch lengthening and if we hope to control the bunch lengthening we should not only care about the inductance but also the loss factor.

Actually, the fact that the resistive part of the impedance can have a strong effect on the longitudinal instability has been pointed out by K. Oide and K. Yokoya in 1990 using an analytical treatment [16]. Also, the longitudinal instability with a purely resistive ring was studied in ref. [17]. Furthermore, ref. [17] shown that reducing the inductance of the SLC damping rings resulted in a lower current threshold according to their measurements, a phenomenon which coincides with our conclusion about the inductance.

Table 2: Beam current threshold estimation with different inductance and longitudinal loss factor

| Inductance $L$ (nH) | Loss factor $k_0$ (V/pc) | Current threshold $I_{th}$ (mA) |
| --- | --- | --- |
| 69 | 1.14 | 5.69 |
| 287 | 1.14 | 23.3 |
| 69 | 3 | 0.88 |

## 4. Conclusion

In this paper we study the longitudinal instability of BEPCII BPR by using Chao-Gareyte scaling law and some challenging theories created by Gao. From the experimental results, we make an estimation that the broad-band impedance varies with frequency as $Z(\omega) \propto \omega^{4.28}$ at lower frequency range and we see that the bunch will lengthen by 30% at the 9.8mA design beam current. The loss factor is calculated to be 1.14V/pc by Gao's theory. Also the total wake potential of the whole ring, bunch lengthening, energy spread increasing and threshold current are estimated analytically. We get the conclusion that the resistive part of the impedance can have strong effects on the longitudinal instability and in order to control the bunch lengthening phenomenon we need to pay more attention to the longitudinal loss factor.

## Acknowledgments

The author thanks the colleagues in accelerator physics group and beam instrumentation group for making the valuable experiments and providing helpful discussions.